\documentclass[twocolumn, twocolappendix,longbib]{aastex7}
\usepackage{amsmath,amssymb,color,bm}

\newcommand{\rev}[1]{#1}

\begin{document}

\title{
Modeling of Collisional Outcomes Based on Impact Simulations of Mars-sized Bodies 
}

\shorttitle{
Modeling of Collisional Outcomes
}

%\correspondingauthor{Hiroshi Kobayashi}

\author{Hiroshi Kobayashi} 
\email{hkobayas@nagoya-u.jp}
\affiliation{Department of Physics, Nagoya University, Nagoya, Aichi 464-8602, Japan}

\author{Hidekazu Tanaka} 
\email{hidekazu@astr.tohoku.ac.jp}
\affiliation{Astronomical Institute, Tohoku University, Aramaki, Aoba-ku, Sendai 980-8578, Japan}

\author{Yukihiko Hasegawa} 
\email{yukihiko.hasegawa@nao.ac.jp}
\affiliation{ALMA Project, National Astronomical Observatory of Japan, 2-21-1 Osawa, Mitaka, Tokyo 181-8588, Japan}

\author{Shu-ichiro Inutsuka} 
\email{inutsuka@nagoya-u.jp}
\affiliation{Department of Physics, Nagoya University, Nagoya, Aichi 464-8602, Japan}

\begin{abstract}

We investigate the outcomes of collisions between Mars-sized bodies through smooth particle hydrodynamics (SPH) simulations, focusing on the transitions among ``merging'', ``hit-and-run'', and catastrophic disruption. By systematically varying impact velocity, angle, and mass ratio, we characterize the dependence of collision outcomes on geometric and energetic parameters. A new analytic model is developed using characteristic energies—particularly the energy deposited in overlapping regions of the colliding bodies—to accurately describe the mass of the largest and second-largest remnants. The model successfully reproduces simulation results across a broad range of impact conditions and improves on previous models by better capturing the transitions between ``merging'', ``hit-and-run'', and disruption. We also derive outcome formulas averaged over impact-parameter-weighted angular distributions, enabling more realistic applications to integrated modeling of planet formation. The model further shows consistency with outcomes from dust aggregate collision simulations, highlighting its utility for modeling collisional processes not only for large planetesimals but also for smaller bodies.

\end{abstract}

\section{Introduction}

Planets are mainly formed via the collisional evolution of small bodies.
Collisions with low velocities result in merging, leading to the mass growth of bodies. Such collisions directly contribute to planet formation.
High-velocity collisions can cause catastrophic disruption, producing substantial ejecta and significant mass loss. If such collisions were the only pathway, an alternative mass-growth mechanism is required for planet formation.
However, some collisions with high angles result in hit-and-run events, where the outcomes closely resemble the original colliders, and only a small amount of ejecta is produced. The contribution of hit-and-run collisions to planet formation depends on their combination with other types of collisions.
Therefore, we construct a model that includes ``hit-and-run'' as well as ``merging'' and catastrophic disruption.

The role of ``hit-and-run'' events is especially important in the late stage of terrestrial planet formation \citep{asphaug06,kokubo10,asphaug21}. For planetesimal accretion, fragmentation can be caused by erosive collisions due to high mass-ratio impacts \citep{kobayashi+10,kobayashi11}. The previous studies carried out such collisional simulations 
%(Benz, Jutzi, Genda, Leinhaldt, Stwart), 
and modeled collisional outcomes \citep{benz99,genda12,jutzi15,leinhardt12,kurosaki23}. 
%(Genda, Leinhaldt, Kokubo). 
In this paper, we revisit this issue. Here, we investigate collisions involving Mars-mass bodies. To reveal the transition between different collisional types, we perform simulations over a range of velocities for several collisional angles. We also explore the dependence on the mass ratio of the colliders. Based on the simulation results, we reconstruct the collisional outcome model, taking into account the overlapping mass of the colliders as determined by the collisional geometry.

In §\ref{sc:simulation}, we present the setup and results of the impact simulations. In §\ref{sc:formula}, we derive the collisional outcome model based on two types of impact energy, normalized by forms of gravitational energy. In §\ref{sc:angle_average}, we calculate the angle-averaged outcomes and provide a semi-analytic formula. In §\ref{sc:discussion}, we discuss our findings and the implications of the new outcome formula for planet formation.

\section{Impact Simulation}
\label{sc:simulation}

\begin{deluxetable*}{c|cccccccccc}
  \tablenum{1}
  \tablecaption{Parameters for Tillotson equation of state \label{tab:tillotson}}
  \tablewidth{0pt}
  \tablehead{
   \colhead{} & \colhead{$\rho_0$[g/cm$^3$]} & \colhead{$a$} & \colhead{$b$} 
  & \colhead{$A$ [g/cm\,s$^2$]} & \colhead{$B$ [g/cm\,s$^2$]} & \colhead{$u_0$ [cm$^2$/s$^2$]} & \colhead{$\alpha_{\rm T}$} 
  & \colhead{$\beta_{\rm T}$} & \colhead{$u_{\rm iv}$ [cm$^2$/s$^2$]} & \colhead{$u_{\rm cv}$ [cm$^2$/s$^2$]} 
  }
% \decimalcolnumbers
 \startdata
  Iron & 7.8 & 0.5 & 1.5 & $1.28\cdot 10^{12}$  & $1.05\cdot 10^{12}$ & $9.5\cdot 10^{10}$ & 5 & 5 & $2.4\cdot 10^{10}$ & $8.7 \cdot 10^{10}$ \\
  Granite & 2.68& 0.5 & 1.3 & $1.8 \cdot 10^{11}$ & $1.8\cdot 10^{11}$ & $1.6\cdot 10^{11}$ & 5 & 5 & $3.5\cdot 10^{10}$ & $1.8 \cdot 10^{11}$\\
  \enddata
 \end{deluxetable*}

We simulate collisions between two bodies using the smooth particle hydrodynamics (SPH): 
The density at the position $\bm x $ is expressed as 
\begin{equation}
 \rho(\bm x) = \sum_i m_i W(\bm x -\bm x_i,h_{i}), 
\end{equation}
where $m_i$, and $\rho_i$
are the mass and density of $i$-th SPH particle, respectively, 
and 
\rev{$h_{i}$ is its smoothing length given by $(m_i/\rho_i)^{1/3}$,}
and $W$ is the kernel function, given by 
\begin{equation}
 W(\bm x,h) = \left(\frac{1}{h\sqrt{\pi}}\right)^3 \exp \left(-\frac{|\bm x|^2}{h^2}\right).
\end{equation}
The equations of the motion and thermal energy for $i$-th SPH particle are given by 
\begin{eqnarray}
 \frac{d \bm v_i}{dt} &=& -\sum_j m_j 
\left(
\frac{P_i}{\rho_i^2}+\frac{P_j}{\rho_j^2} + \Pi_{ij}
\right) \frac{\partial}{\partial \bm x_i} W(\bm x_i-\bm x_j,h_{ij})
\label{eq:eom} 
\nonumber
\\
 && - G \sum_j \hat m_j \frac{\bm x_i- \bm x_j}{|\bm x_i- \bm x_j|^3},\\
 \frac{d u_i}{dt} &=& \frac{1}{2}\sum_j m_j 
\left( \frac{P_i}{\rho_i^2}+\frac{P_j}{\rho_j^2} + \Pi_{ij} \right) 
\nonumber\label{eq:energy_equation} 
\\
&& \times 
(\bm v_i-\bm v_j) \cdot \frac{\partial}{\partial \bm x_i} W(\bm x_i-\bm x_j,h_{ij}), 
\end{eqnarray}
where $P_i$ and $u_i$ are the pressure and internal energy of the $i$-th particle, respectively, $\Pi_{ij}$ is the artificial viscosity, $G$ is the gravitational constant, $h_{ij}$ is the average smoothing length of $i$-th and $j$-th SPH particles, and $\hat m_j$ is the effective mass of $j$-th particle defined by 
\begin{equation}
 \hat m_j = m_j  \int_0^{|\bm x_i - \bm x_j|} 4 \pi r^2 W(r,h_{ij}) dr. 
\end{equation}
We approximate $\hat m_j = m_j$ if $|\bm x_i - \bm x_j| \geq 2 h_{ij}$. 
The artificial viscosity is given by \citep{monaghan92}
\begin{equation}
 \Pi_{ij} = 
\left\{
\begin{array}{l}
- \alpha_{\rm vis} \frac{c_{{\rm s},ij}}{\rho_{ij}}  - \beta_{\rm vis} \frac{\Xi_{ij}^2}{\rho_{\rm ij}}, \\ 
\quad \quad \quad {\rm if} (\bm x_i-\bm x_j)\cdot (\bm v_i - \bm v_j) < 0\\
0 \quad \quad {\rm otherwise}, 
\end{array}
\right.
\end{equation}
where $c_{{\rm s},ij}$ and $\rho_{ij}$ are the average sound speed and density of the $i$-th and $j$-th particles, respectively, and 
\begin{equation}
 \Xi_{ij} = \frac{h_{ij} (\bm x_i-\bm x_j)\cdot (\bm v_i - \bm v_j)}{|\bm x_i-\bm x_j|^2+0.01h_{ij}^2}. 
\end{equation}
We set $\alpha_{\rm vis} = 1.5$ and $\beta_{\rm vis} = 3.0$ according to 
\citet{monaghan92}. 

\begin{figure*}[hbt]
%\plotone{snapshots.pdf}
\plotone{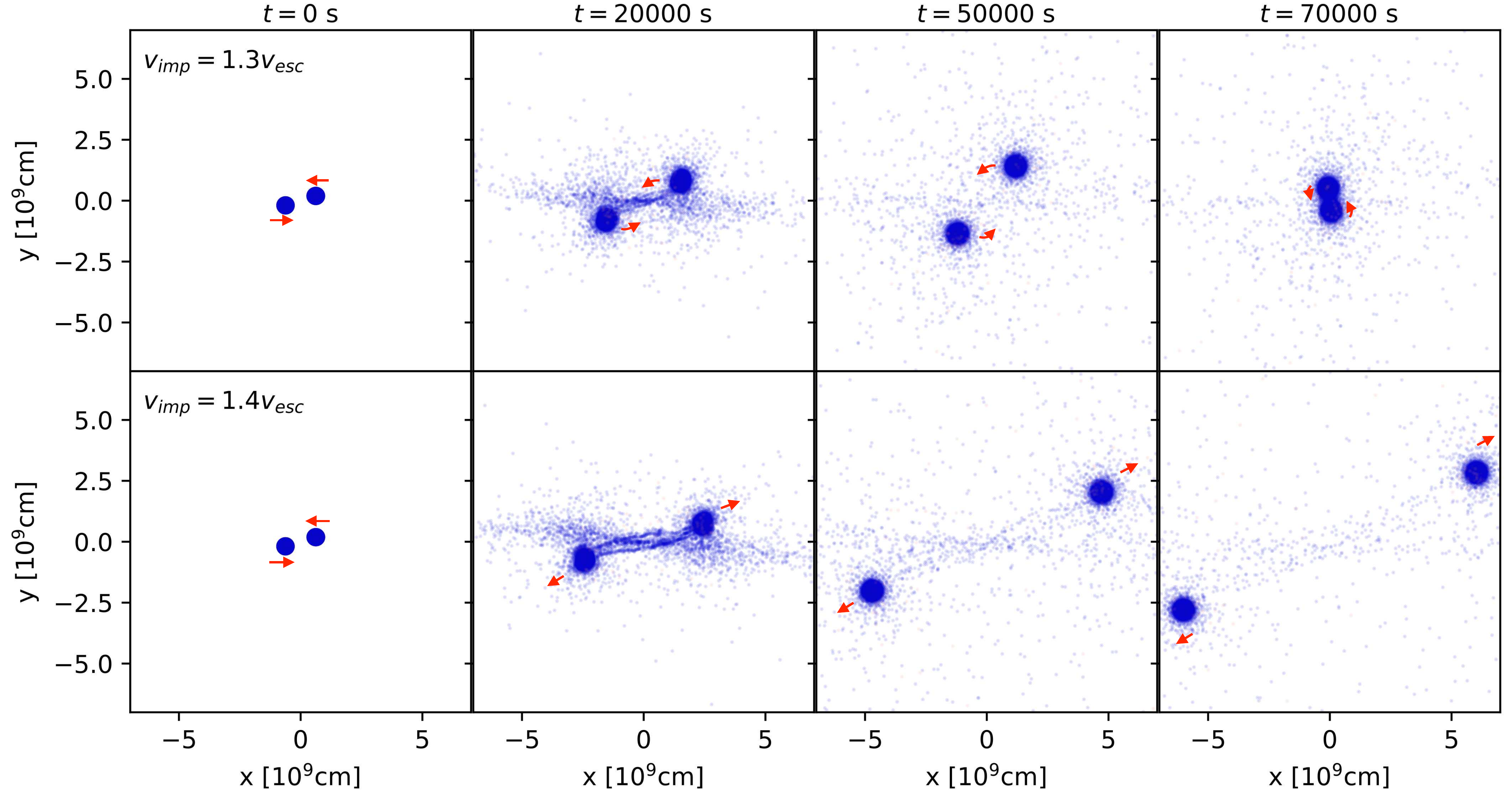}
%\epsscale{0.9}
\figcaption{Snapshots of impacts between Mars-mass bodies with $\theta = 30^\circ$ and $v_{\rm imp}/v_{\rm esc} =1.3$ (top) and 1.4 (bottom). Red arrows roughly indicate the overall motion of two major objects in each snapshot.
\label{fig:snap_shots}} 
\end{figure*}

We adopt the Tillotson equation of state for iron and granite \citep{melosh89}, and treat colliding bodies with iron cores (30 wt.\%) and granite mantles (70 wt.\%). 
As we analytically show in \S~\ref{sc:formula}, the choice of detailed compositions of core and mantle is not important for collisional outcomes of Mars-size or larger target bodies.  
In the Tillotson equation state, pressure is written by the parameters listed in Table \ref{tab:tillotson}. In the solid state, pressure is given by 
\begin{equation}
 P = \left[a + \frac{b}{u \rho_0^2 /u_0 \rho^2 +1} \right] \rho u + A \frac{\rho - \rho_0}{\rho_0} + B
\left(
\frac{\rho - \rho_0}{\rho_0} 
\right)^2, \label{eq:tillotson}
\end{equation}
where $u$ is the internal energy, and $a,b,A,B,\rho_0$ and $u_0$ are the Tillotson parameters. In the expansion state with high internal energy, pressure is determined by vaporization, which is expressed by other forms (see details in Appendix \ref{sc:tillotson}). 

We perform simulations by time integration of Eqs.~(\ref{eq:eom}) and (\ref{eq:energy_equation}) 
using parallel supercomputers with effective parallelization by Framework for Developing Particle Simulator
\citep[FDPS,][]{iwasawa16}. 
%(FDPS, Iwasawa et al. 2015, 2016). 
The self-gravity is calculated by the tree algorithm in FDPS \citep{iwasawa16}. 

\begin{table}[b]
 \caption{Parameters of impact simulations \label{tab:parameter}}
\begin{tabular}{c | c}
%  \tablenum{2}
%  \tablewidth{0pt}
% \tablehead{none}
% \tablehead{\colhead{}}
% \decimalcolnumbers
%  \startdata
\hline \hline
 $M_{\rm tar}/M_{\rm pro}$ & 1, 3, 10, and 30 \\
 $\theta$ & $30^\circ, 45^\circ$, and $60^\circ$ \\
 $v_{\rm imp}/v_{\rm esc}$ & 1.1, 1.2, 1.3, 1.5, 2, 4, 8, and 16\\
\hline
%  \enddata
 \end{tabular}
\end{table}

We consider collisions between target and projectile bodies with masses $M_{\rm tar}$ and $M_{\rm pro}$, respectively. We set $M_{\rm tar} = 6 \times 10^{26}$\,g using 36000 SPH particles. 
The collisional parameters of our simulations are the mass ratio $M_{\rm tar}/M_{\rm pro}$, the collisional angle $\theta$ (head-on collisions for $\theta = 0^\circ$), and $v_{\rm imp}/v_{\rm esc}$, where $v_{\rm imp}$ is the collisional velocity and $v_{\rm esc}$ is the mutual escape velocity of colliding bodies. Note that $v_{\rm imp}$ is the velocity at the time of collision, which is different from the initial velocities in simulations. We set the initial velocities for simulations according to the solution of the two body problem with $v_{\rm imp}$ and the initial separations of $2(R_{\rm tar}+R_{\rm pro})$, where $R_{\rm tar}$ and $R_{\rm pro}$ are the radii of target and projectile, respectively. 
We carry out simulations for 96 parameter sets listed in Table \ref{tab:parameter}. 

Figure \ref{fig:snap_shots} shows snapshots of collisions with $\theta = 30^\circ$, $M_{\rm tar} = M_{\rm pro}$, and $v_{\rm imp}/v_{\rm esc} = 1.3$ (top) and 1.4 (bottom). In the panels at $t = 0$\,s, the upper bodies move to the left, while the lower bodies move to the right.  The collisions result in two large remnants originating from the colliding bodies at $t= 2 \times 10^4$\,s. For $v_{\rm imp}/v_{\rm esc} = 1.3$, the remnants collide with each other again at $t = 7 \times 10^4$\,s, leading to a large merger. This type of collision is called ``merging''. For $v_{\rm imp}/v_{\rm esc} = 1.4$, the remnants fly apart, resulting in two similarly large remnants. This type of collision is called ``hit-and-run''. Therefore, a small difference in $v_{\rm imp}$ leads to significantly different outcomes. 

To analyze collisional outcomes, we measure the masses of largest and second largest bodies resulting from a collision, $M_{\rm lar}$ and $M_{\rm s}$, respectively. These bodies are identified via the ``Friend-of-Friend'' algorithm, including the gravitational binding \citep{benz99}. The calculated values of $M_{\rm lar}$ and $M_{\rm s}$ are almost fixed by $t=10^5$\,s and we safely analyze the data at $t=2\times 10^5$\,s. 
The measurement of the second largest remnant mass, $M_{\rm s}$, is required for the examination of ``hit-and-run'' and ``merging'', while the largest remnant mass $M_{\rm lar}$ is important to recognize ``catastrophic disruption''. 

\begin{figure*}[hbt]
\plotone{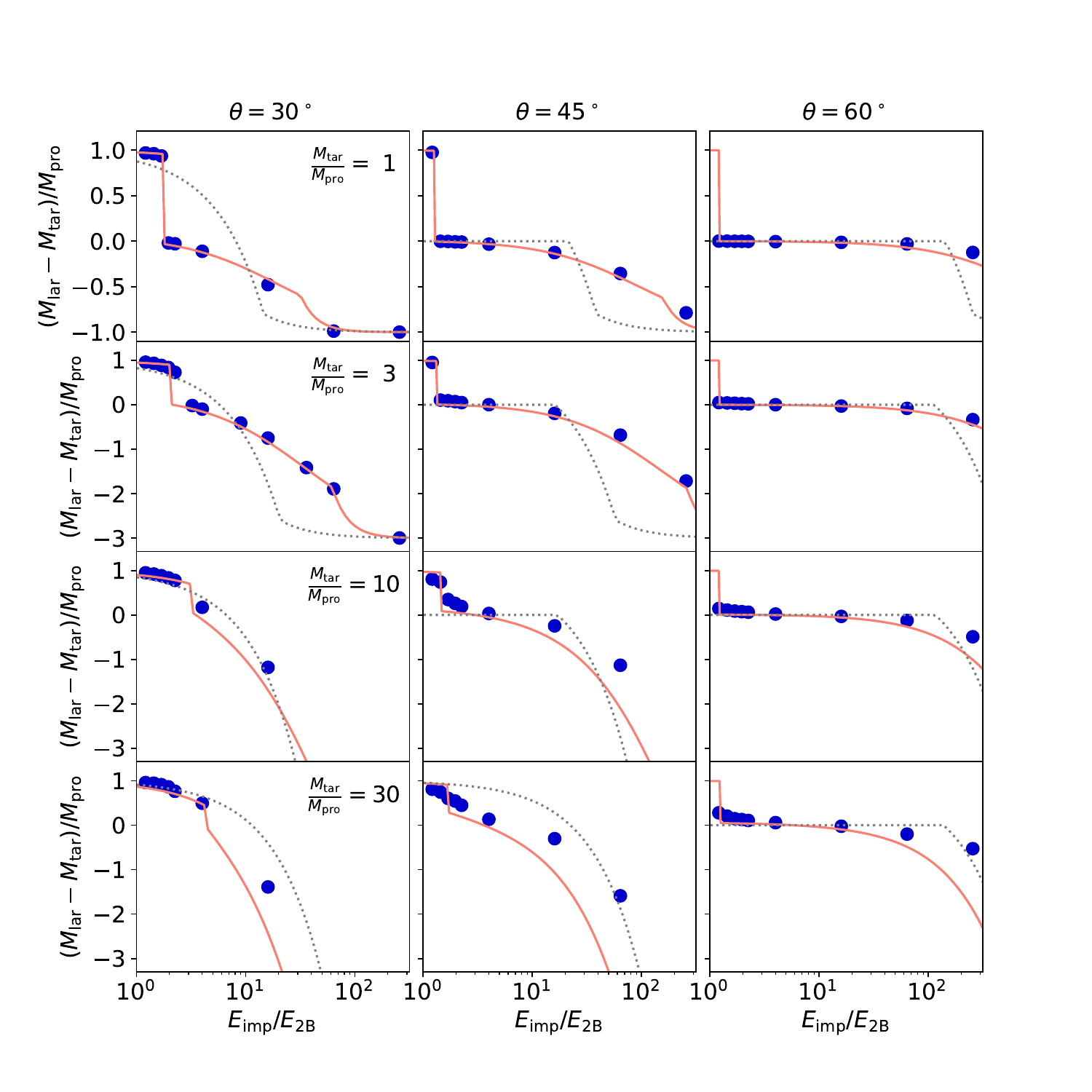}
%\plottwo{fig_t100d1e-3_long.pdf}{fig_d1e-3_mp_sigma_port.pdf}
\figcaption{
Mass of the largest post-collision remnant, $M_{\rm lar}$, is shown as a form
$(M_{\rm lar}-M_{\rm tar})/M_{\rm pro}$ to illustrate growth or erosion relative to the target mass $M_{\rm tar}$ and the projectile mass $M_{\rm pro}$, as a function of the ratio of impact energy $E_{\rm imp}$ to the two-body gravitational energy $E_{\rm 2B}$. Blue dots represent the results of simulations. Red curves indicate the analytic formula given by Eq.~(\ref{eq:ana_mlar}). 
\rev{Crosses denote the transition between gravity-dominated (low energy) and vaporization-dominated (high energy) regimes. }
We find $(M_{\rm lar}-M_{\rm tar})/M_{\rm pro} \sim 1$ if $E_{\rm res} \la E_{\rm 2B}$ (details are described in the text). 
Dotted curves show the model by \citet{leinhardt12}. 
%modified by target mass $M_{\rm tar}$ and projectile mass $M_{\rm pro}$, 
\label{fig:comp_ml}
} 
\end{figure*}

\begin{figure*}[hbt]
\plotone{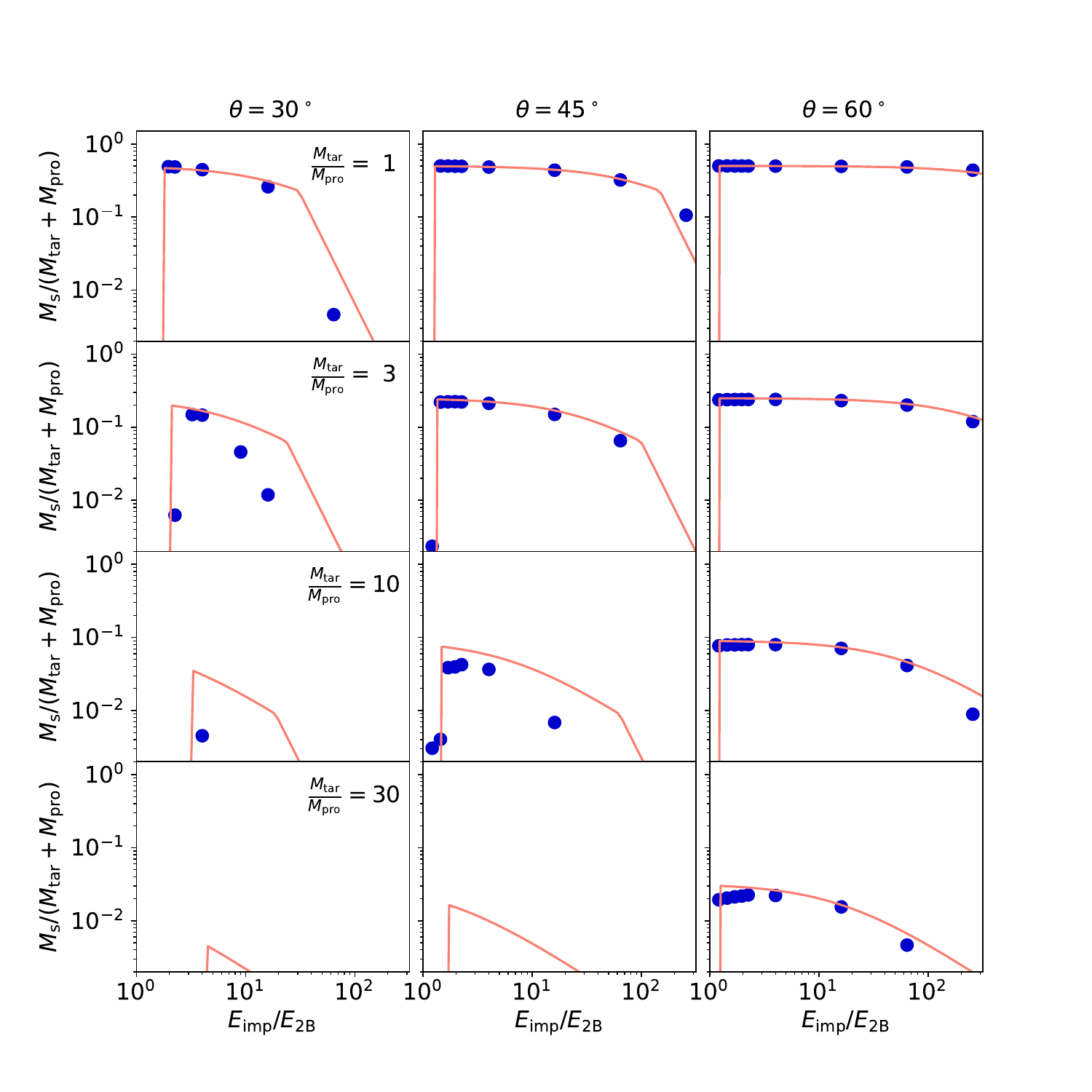}
%\plottwo{fig_t100d1e-3_long.pdf}{fig_d1e-3_mp_sigma_port.pdf}
\figcaption{
Mass of the second-largest remnant, $M_{\rm s}$ normalized by the total mass of colliding bodies, $M_{\rm tar}+M_{\rm pro}$. 
Blue dots represent the results of simulations. Red curves indicate the analytic formula given by Eq.~(\ref{eq:ana_ms}). \rev{Crosses denote the transition between gravity-dominated (low energy) and vaporization-dominated (high energy) regimes. }
For $\theta = 30^\circ$ and $45^\circ$ with $M_{\rm tar} = 30 M_{\rm pro}$, $M_{\rm s}$ resulting from simulations are smaller than $2 \times 10^{-3} (M_{\rm tar}+M_{\rm pro})$, and hence no simulation data is plotted. 
\label{fig:comp_second}
} 
\end{figure*}

\begin{figure*}[hbt]
\plotone{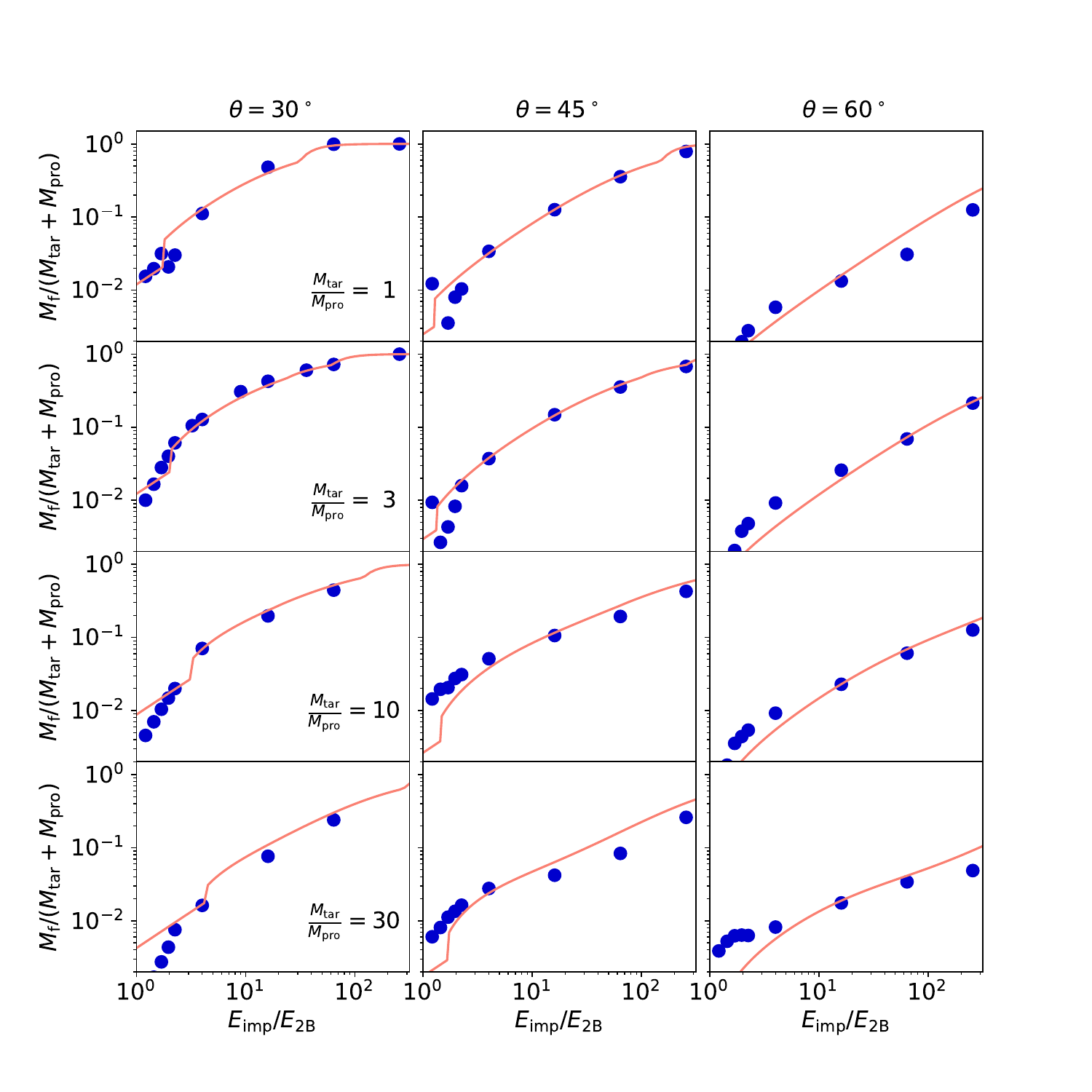}
%\plottwo{fig_t100d1e-3_long.pdf}{fig_d1e-3_mp_sigma_port.pdf}
\figcaption{Total mass of remnants smaller than the second-largest remnant, $M_{\rm f}$, normalized by the total mass of colliding bodies, $M_{\rm tar}+M_{\rm pro}$. Blue dot represent the results of simulations. Red curves indicate the analytic formula given by $M_{\rm f} = M_{\rm tar}+M_{\rm pro}-M_{\rm lar}-M_{\rm s}$ with Eqs.~(\ref{eq:ana_mlar}) and (\ref{eq:ana_ms}). 
\label{fig:comp_frag}
} 
\end{figure*}

We investigate $M_{\rm lar}$, $M_{\rm s}$, and fragment masses $M_{\rm f} = M_{\rm tar}+M_{\rm pro} -M_{\rm lar} - M_{\rm s}$, as a function of $E_{\rm imp}/E_{\rm 2B} = (v_{\rm imp}/v_{\rm esc})^2$, where 
$E_{\rm imp}$ is the impact energy and $E_{\rm 2B}$ is the two-body gravitational energy required to separate two bodies in contact to infinite distance. 
They are given by
\begin{eqnarray}
  E_{\rm imp} &=& \frac{1}{2} \frac{M_{\rm tar} M_{\rm pro}}{M_{\rm tar}+M_{\rm pro}} v_{\rm imp}^2,\label{eq:e_imp}\\
 E_{\rm 2B} &=& \frac{G M_{\rm tar}M_{\rm pro}}{R_{\rm tar}+R_{\rm pro}},
\nonumber\label{eq:e_2b}
\\
&=& \frac{1}{2} \frac{M_{\rm tar} M_{\rm pro}}{M_{\rm tar}+M_{\rm pro}} v_{\rm esc}^2.
\end{eqnarray}
Figures \ref{fig:comp_ml} and \ref{fig:comp_second} show $(M_{\rm lar} - M_{\rm tar})/M_{\rm pro}$ and $M_{\rm s}/(M_{\rm tar}+M_{\rm pro})$, respectively. 
For $M_{\rm tar} /M_{\rm pro} = 1$ and 3, $(M_{\rm lar} - M_{\rm tar})/M_{\rm pro} \sim 0$ and $M_{\rm s}/(M_{\rm tar}+M_{\rm pro}) \sim M_{\rm pro}/(M_{\rm tar}+M_{\rm pro})$ at $E_{\rm imp}/E_{\rm 2B} \sim 10$. These results indicate $M_{\rm lar} \sim M_{\rm tar}$ and $M_{\rm s} \sim M_{\rm pro}$, which are caused by ``hit-and-run'' collisions. On the other hand, $(M_{\rm lar} - M_{\rm tar})/M_{\rm pro} \sim 1$ at $E_{\rm imp}/E_{\rm 2B} \sim 1$, which are caused by ``merging''. 
The ``merging'' collisions result in $M_{\rm lar} \sim M_{\rm tar} + M_{\rm pro}$ and $M_{\rm s} \ll M_{\rm pro}$. For high energy collisions with $E_{\rm imp}/E_{\rm 2B} \ga 10$, $M_{\rm f}$ smoothly increases with impact energies, because of  ``erosive'' collisions (see Figure~\ref{fig:comp_frag}). 

%The collisional outcomes are not characterized as ``merging'' or ``disruption''. 

The collisional outcomes depend on $\theta$, which is caused by different transitional energies from ``merging'' to ``hit-and-run'' collisions and from ``hit-and-run'' to ``erosive'' collisions. 
The transitional energy from ``merging'' to ``hit-and-run'' collisions is easily recognized as the energy where $(M_{\rm lar}- M_{\rm tar})/M_{\rm pro}$ falls from 1 to 0 in Figure \ref{fig:comp_ml}. 
The transition energy tends to be higher for small $\theta$. This is explained by the large energy dissipation for low $\theta$, because geometrically overlapping volumes in colliders are significant. 
On the other hand, ``erosive'' collisions are characterized by significant ejecta mass compared to $M_{\rm pro}$. The transition energy to ``erosive'' collisions is thus estimated from 
$(M_{\rm lar}- M_{\rm tar})/M_{\rm pro} \approx 1$, as shown in Figure~\ref{fig:comp_ml}. The transition energy is low for small $\theta$ (see Figure \ref{fig:comp_ml}). Shock heating from an impact induces the ejection of fragments from colliding bodies \citep{genda15a}. For small $\theta$, shock heating energies are large because of large overlapping volumes, making collisional erosion particularly effective. 

%For low $\theta$, ``merging'' collisions occurs

\section{Collisional outcome formula}
\label{sc:formula}

\begin{figure}
 \plotone{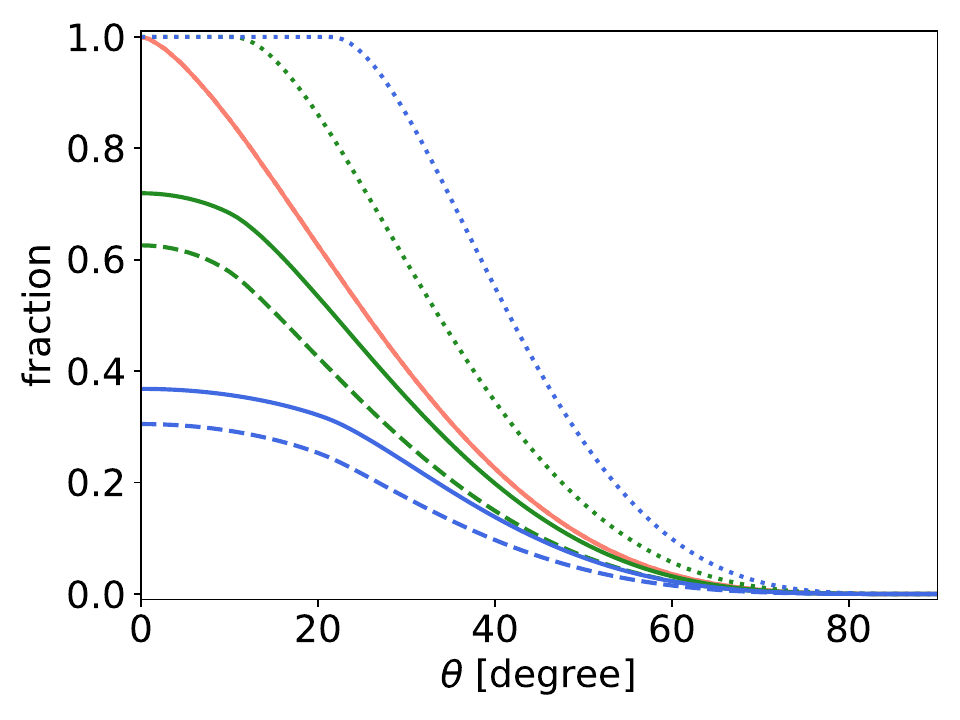}
 \figcaption{\rev{$M_{\rm t,o}/M_{\rm tar}$ (dashed line), $M_{\rm p,o}/M_{\rm pro}$ (dotted line), $\gamma^\prime$ (solid line) obtained from the formula in Appendix~\ref{sc:overlap} for $M_{\rm tar} / M_{\rm pro} = 1$ (red), 3 (green), 10 (blue). The three lines coincide for $M_{\rm tar} / M_{\rm pro} = 1$. }
\label{fig:overlap_fraction}}
\end{figure}

In an impact between bodies, the overlap of their volumes plays a critical role in transferring energy and momentum. We calculate the overlapping masses in target and projectile bodies, $M_{\rm t,o}$ and $M_{\rm p,o}$, respectively. This is done with the formulae provided in Appendix~\ref{sc:overlap}, employing the masses and radii of cores and mantles. \rev{Figure~\ref{fig:overlap_fraction} shows the impact-angle dependence of $M_{\rm t,o}$ and $M_{\rm p,o}$ obtained from the formulae in Appendix~\ref{sc:overlap}.}

The impact energy contained within these overlapping masses primarily facilitates the exchange of energy and momentum between the colliding bodies. The velocities of target and projectile with respect to the center of gravity are, respectively, given by 
\begin{equation}
     \frac{ M_{\rm pro} v_{\rm imp} }{M_{\rm tar}+M_{\rm pro}}, \quad 
      \frac{M_{\rm tar}v_{\rm imp}}{M_{\rm tar}+M_{\rm pro}}.
\end{equation}
The total impact energy of the overlapping masses is thus calculated as
\begin{eqnarray}
  E_{\rm over} 
   &=& \frac{1}{2} 
    M_{\rm t,o}
    \left(
     \frac{ M_{\rm pro} v_{\rm imp} }{M_{\rm tar}+M_{\rm pro}}\right)^2
    \nonumber \\
   && + \frac{1}{2} M_{\rm p,o}
   \left(\frac{M_{\rm tar}v_{\rm imp}}{M_{\rm tar}+M_{\rm pro}}\right)^2,\\
  &=& \gamma^{\prime} E_{\rm imp},
\end{eqnarray}
where the overlapping energy fraction $\gamma^\prime$ is given by 
\begin{eqnarray}
\gamma^\prime &=& \frac{1}{ M_{\rm tar}+M_{\rm pro}} \left(\frac{M_{\rm t,o}}{M_{\rm tar}} M_{\rm pro} +  \frac{M_{\rm p,o}}{M_{\rm pro}}M_{\rm tar}\right).
\end{eqnarray}
\rev{The dependence of $\gamma^\prime$ on $\theta$ is shown in Figure~\ref{fig:overlap_fraction}.}

However, to be more precise, the component of kinetic energy due to impact velocity perpendicular to the surfaces mainly converts into the thermal energy by the shock wave generated by the impact. The outcome is thus characterized by the perpendicular overlapping impact energy, given by 
\begin{eqnarray}
 E_{\rm over,\perp} &=&   E_{\rm over} \cos^2 \theta,\label{eq:eoperp} \\
&=& \gamma^{\prime} E_{\rm imp} \cos^2 \theta, 
\end{eqnarray}
%and $v_{\rm imp}$ and $\theta$ are the impact velocity and angle , respectively. 
%$E_{\rm i}$ is the total impact energy. 

When a collision occurs, a shock wave converts the kinetic energy into the thermal energy and significantly modifies the velocities. The shock removes $\sim E_{\rm over,\perp}$ from the kinetic energy of the colliding bodies, and then the two major bodies originate from the colliders only have their orbital energy of $\sim E_{\rm res}$, where 
\begin{equation}
 E_{\rm res} = E_{\rm imp} - E_{\rm over,\perp}. 
\end{equation}
If the residual kinetic energy, $E_{\rm res}$, is much smaller than the two-body gravitational energy $E_{\rm 2B}$, the two major bodies merge under the influence of their mutual gravity.
On the other hand, large $E_{\rm over,\perp}$ induces the ejection of high-velocity fragments from the vicinity of the impact point. If $E_{\rm over,\perp}$ is much larger than the gravitational energy of the target, the projectile, or a merger, its mass is reduced by the ejection. 
For the target, the projectile, and the merger resulting from perfect accretion, the gravitational energies required to disperse uniform spheres to infinity 
are, respectively, given by 
\begin{eqnarray}
 E_{\rm g,tar} &=& \frac{3 G M_{\rm tar}^2}{5R_{\rm tar}},\\
 E_{\rm g,pro} &=& \frac{3 G M_{\rm pro}^2}{5R_{\rm pro}},\\
 E_{\rm g,tot} &=& \frac{3 G (M_{\rm tar}+M_{\rm pro})^2}{5R_{\rm tot}},
\end{eqnarray}
where $R_{\rm tot}$ is the radius of bodies with the mass of $M_{\rm tar}+M_{\rm pro}$. We here approximately assume $R_{\rm tot} = R_{\rm tar} (1+M_{\rm pro}/M_{\rm tar})^{1/3}$. 
We formulate a collisional outcome model using these characteristic energies. 

%{\bf Largest remnant}:
We first provide an analytic formula for the mass of the largest remnant, $M_{\rm lar}$, expressed as a function of the following dimensionless energies: 
\begin{eqnarray}
%\tilde{E}_{\rm g,r} &=& E_{\rm res}/E_{\rm 2B},\\
\tilde{E}_{\rm g,tot} &=& E_{\rm g,tot}/E_{\rm over,\perp}, 
\label{eq:til_egtot}
\\
\tilde{E}_{\rm g,tar} &=& E_{\rm g,tar}/E_{\rm over,\perp},
\label{eq:til_egtar} 
 \\
\tilde{E}_{\rm v,tot} &=& [(M_{\rm tar}+M_{\rm pro}) u_{\rm cv}]/E_{\rm over,\perp},\label{eq:til_evtot}
\\ 
\tilde{E}_{\rm v,tar} &=& (M_{\rm tar} u_{\rm cv})/E_{\rm over,\perp}, 
\label{eq:til_evtar} 
\end{eqnarray}
where $u_{\rm cv}$ represents the specific energy needed for complete vaporization (see Table \ref{tab:tillotson}). For simplicity, $u_{\rm cv}$ is calculated from the value of a mantle material. 

As discussed above, the mass loss due to collisional erosion or catastrophic disruption is controlled by $\tilde{E}_{\rm g,tot}$ for the merged body and $\tilde{E}_{\rm g,tar}$ for the hit-and-run remnant originating from the target. Meanwhile the mass loss due to collisional vaporization is characterized by $\tilde{E}_{\rm v,tot}$ for the merged body and $\tilde{E}_{\rm v,tar}$ for the hit-and-run remnant originating from the target. In addition, $E_{\rm res} \sim E_{\rm 2B}$ marks the transition between ``merging'' and ``hit-and-run'' regimes. 

For a fixed $\theta$, $M_{\rm lar}$ is inversely proportional to $E_{\rm imp}$ for high $E_{\rm imp}$, while $M_{\rm lar} \sim M_{\rm tar}+M_{\rm pro}$ or $\sim M_{\rm tar}$ in ``merging'' or ``hit-and-run'' regimes for low $E_{\rm imp}$ 
\citep{genda17,suetsugu18}. 
As discussed above, $M_{\rm lar}$ is determined by a certain function of the dimensionless energies in Eqs.~(\ref{eq:til_egtot}) or (\ref{eq:til_egtar}) unless vaporization occurs. We approximately model this dependence by the following simple function, 
\begin{equation}
  F_{\rm g}(x) = 1- (1 + x / 0.15)^{-1}, 
\end{equation}
where $\tilde{E}_{\rm g,tot}$ or $\tilde{E}_{\rm g,tar}$ is substituted for $x$. 
$F_{\rm g}(x) \propto x$ for $x \ll 1$ and $F_{\rm g}(x) \sim 1$ for $x \gg 1$. Because $\tilde{E}_{\rm g,tot}$ and $\tilde{E}_{\rm g,tar}$ are inversely proportional to $E_{\rm imp}$, $F_{\rm g}(\tilde{E}_{\rm g,tot})$ and $F_{\rm g}(\tilde{E}_{\rm g,tar})$ conveniently show the dependence of $M_{\rm lar}$ on $E_{\rm imp}$. 
For ``merging'', $M_{\rm lar}$ is determined by 
\begin{equation}
  M_{\rm g,tot} = (M_{\rm tar} + M_{\rm pro}) F_{\rm g}(\tilde{E}_{\rm g,tot}).
\label{eq:mgtot} 
\end{equation}
For ``hit-and-run'', $M_{\rm lar}$ is given by 
\begin{equation}
 M_{\rm g,tar} = M_{\rm tar} F_{\rm g}(\tilde{E}_{\rm g,tar}) + M_{\rm s,gain},
\label{eq:mgtar} 
\end{equation}
where $M_{\rm s,grain}$, the mass transferred from the projectile to target bodies, is defined below in Eq.~(\ref{eq:msgain}). 
%Then $M_{\rm lar}$ is given by $(M_{\rm tar}+M_{\rm pro}) F_{\rm g}(\tilde{E}_{\rm g,tot})$ or $M_{\rm tar} F_{\rm g}(\tilde{E}_{\rm g,tar})$. 
On the other hand, extremely high $E_{\rm imp}$ induces vaporization. 
As discussed above, if vaporization is dominant, $M_{\rm lar}$ is determined by the dimensionless energy of Eqs.~(\ref{eq:til_evtot}) or (\ref{eq:til_evtar}). 
Vaporization is effective if $E_{\rm imp}$ exceeds the vaporization threshold. 
The total mass of impact vapors is proportional to $E_{\rm imp}$ if $E_{\rm imp}$ is much larger than the threshold vaporization energy \citep{melosh89}. However, the vaporized mass by an impact has a stronger dependence on $E_{\rm imp}$ if $E_{\rm imp}$ is only several times larger than the vaporization threshold \citep{miyayama24}. At the transition from the mass loss due to dynamical ejection to that by vaporization, such a sharp dependence on vaporization is important. \rev{In addition, only about half of $E_{\rm over,\perp}$ is converted into thermal energy by shock and hence vaporization becomes effective when ${\tilde E}_{\rm v,tot} \ga 1/2$ or ${\tilde E}_{\rm v,tar} \ga 1/2$.} We thus model the vaporization function using a cubic function of dimensionless energy, such as 
\begin{equation}
  G_{\rm v}(x) = {\rm Min}[(2.5x)^{3},1],  
\end{equation}
where $\tilde{E}_{\rm v,tot}$ or $\tilde{E}_{\rm v,tar}$ is substituted for $x$. Our choice of this simple expression is convenient enough for describing the simulation results. 
%we assume the cubic function $G_{\rm v}$ is used in the similar manner to $F_{\rm g}$. 
If vaporization is effective,  
$M_{\rm lar}$ 
for ``merging'' and ``hit-and-run'' are determined by 
\begin{eqnarray}
 M_{\rm v,tot} &=& (M_{\rm tar} + M_{\rm pro}) G_{\rm v}(\tilde{E}_{\rm v,tot}),
\label{eq:mvtot} \\
 M_{\rm v,tar} &=& M_{\rm tar} G_{\rm v}(\tilde{E}_{\rm v,tar}). 
\label{eq:mvtar}
\end{eqnarray}
\rev{It should be noted that the vaporization-dominated regime is limited only for high velocity and low angle (see Figures \ref{fig:comp_ml} and \ref{fig:comp_second}).}

Using Eqs~(\ref{eq:mgtot}),
(\ref{eq:mgtar}),
(\ref{eq:mvtot}), and (\ref{eq:mvtar}), 
we provide $M_{\rm lar}$ as follows: 
\begin{equation}
 M_{\rm lar} = \left\{
  \begin{array}{l}
   {\rm Min}(M_{\rm g,tot},M_{\rm v,tot}) \,\, {\rm for}\, E_{\rm res} \leq 1.2 E_{\rm 2B}, \\
%\\  
%& {\rm for}\, \tilde{E}_{\rm g,r} \leq 1.2, \\
%\quad \quad \quad \quad \quad {\rm for}\, E_{\rm res} \leq 1.2 E_{\rm 2B}, \\
   {\rm Min}(M_{\rm g,tar},M_{\rm v,tar}) \,\, {\rm for}\, E_{\rm res} > 1.2 E_{\rm 2B}, 
%\\ \quad \quad \quad 
%&{\rm for}\, \tilde{E}_{\rm g,r} > 1.2,
%\\ \quad \quad \quad \quad \quad {\rm for}\, E_{\rm res} > 1.2 E_{\rm 2B}, 
  \end{array}
 \right.\label{eq:ana_mlar}
 \end{equation}
where the function ${\rm Min}(x,y)$ results in the smaller of $x$ and $y$. 

%$\tilde{M}_{\rm s,grain}$ is the mass transfer from the projectile to target bodies, which will be defined below. 

%{\bf Second largest body:}
The mass of the second largest body, $M_{\rm s}$, is determined by the similar manner to $M_{\rm lar}$. We additionally introduce dimensionless energies:
\begin{eqnarray}
 \tilde{E}_{\rm g,pro} &=& E_{\rm g,pro}/E_{\rm over,\perp}, \\
\tilde{E}_{\rm v,pro} &=& M_{\rm pro} u_{\rm cv}/E_{\rm over,\perp}. 
\end{eqnarray}
The mass loss of the hit-and-run remnant originating from the projectile is characterized by $\tilde{E}_{\rm g,p}$ for collisional erosion and $\tilde{E}_{\rm v,p}$ for collisional vaporization. 

For ``hit-and-run'', $M_{\rm s}$ is modeled similarly to $M_{\rm lar}$; 
$M_{\rm s}$ is determined by $F_{\rm g}$ or $G_{\rm v}$ for dynamical ejection or vaporization, respectively. We provide $M_{\rm s}$ as 
\begin{eqnarray}
  M_{\rm g,pro} &=& M_{\rm pro} F_{\rm g}(\tilde{E}_{\rm g,pro}), 
%1-[1 + (0.15 \tilde{E}_{\rm g,p})^{-1}]^{-1}, 
  \label{eq:ms_loss} 
\\
 M_{\rm v,pro} &=& M_{\rm pro} G_{\rm v}(\tilde{E}_{\rm v,pro}).
%{\rm Min}[(\tilde{E}_{\rm v,p}/2.5)^{-3},1].
\end{eqnarray}
For ``merging'', 
$M_{\rm s}$ is much smaller than the total ejection mass from the largest remnant, $M_{\rm tar} + M_{\rm pro} - M_{\rm lar}$. We model $M_{\rm s}$ as 
\begin{equation}
  M_{\rm g,tot,s} = 0.01 (M_{\rm tar} + M_{\rm pro} - M_{\rm g,tot}) . 
\end{equation}
We then provide $M_{\rm s}$ as 
\begin{equation}
 M_{\rm s} = \left\{
\begin{array}{c c}
 M_{\rm g,tot,s} & {\rm for}\, E_{\rm res} \leq 1.2 E_{\rm 2B},\\
 {\rm Min}(M_{\rm g,pro},M_{\rm v,pro}) & {\rm for}\, E_{\rm res} > 1.2 E_{\rm 2B}. 
\end{array}
\right.\label{eq:ana_ms}
\end{equation}
%where

The ejected mass from the projectile can be bound by the gravity of the largest remnant. The kinetic energy of ejecta from the projectile is roughly proportional to $E_{\rm res}$. We model the transferred mass from $M_{\rm pro}$ to $M_{\rm tar}$ using the ejecta mass from $M_{\rm pro}$ and $E_{\rm res}/E_{\rm 2B}$, such as 
\begin{eqnarray}
 M_{\rm s,gain} &=& 
\left(M_{\rm pro} - M_{\rm g,pro} \right) 
\frac{E_{\rm 2B}}{E_{\rm res}}. 
\label{eq:msgain}
%\frac{[1 + (0.15 \tilde{E}_{\rm g,p})^{-1}]^{-1} M_{\rm pro}}{ \tilde{E}_{\rm g,r} M_{\rm tar} }.
\end{eqnarray}

The red curves in Figures \ref{fig:comp_ml}--\ref{fig:comp_frag} show the analytical formulae for $M_{\rm lar}$ and $M_{\rm s}$. The analytic formulae well reproduce the collisional outcomes obtained from simulations. \citet{leinhardt12} provided the similar formulae for $M_{\rm lar}$, which are plotted as dotted curves in Figure \ref{fig:comp_ml}. From the comparison of the formulae, our formula agrees with the results by simulations more. Especially, the transition from ``merging'' to ``hit-and-run'' collisions are reproduced better. 

It should be noted that the formula for $M_{\rm s}$ is in good agreement with the results of simulations only for $M_{\rm tar}/M_{\rm pro} \leq 3$ or $\theta = 60^\circ$. Even for the parameters, ``merging'' collisions result in small $M_{\rm s}$, which are not reproduced well by the analytic formula. 
In addition, for $M_{\rm tar}/M_{\rm pro} > 10$ and $\theta > 45^\circ$, $M_{\rm s}$ is much smaller than $M_{\rm tar} + M_{\rm pro}$. For such low $\theta$ and small $M_{\rm pro}$, $M_{\rm s}$ may be determined by extra effects that we do not consider above. However, this effect is minor for the collisional outcome model because second largest remnants are as small as other fragments ($M_{\rm s} \sim M_{\rm f}$). 

Although $M_{\rm f}$ is given by the analytic formulae of $M_{\rm lar}$ and $M_{\rm s}$, the analytic formula is good agreement with the simulation results for $M_{\rm f}$ (Figure \ref{fig:comp_frag}). Although the disagreement of $M_{\rm s}$ influences $M_{\rm f}$, our model explains the energy dependence of $M_{\rm f}$. 

\section{average outcomes}
\label{sc:angle_average}

\begin{figure*}[hbt]
\plottwo{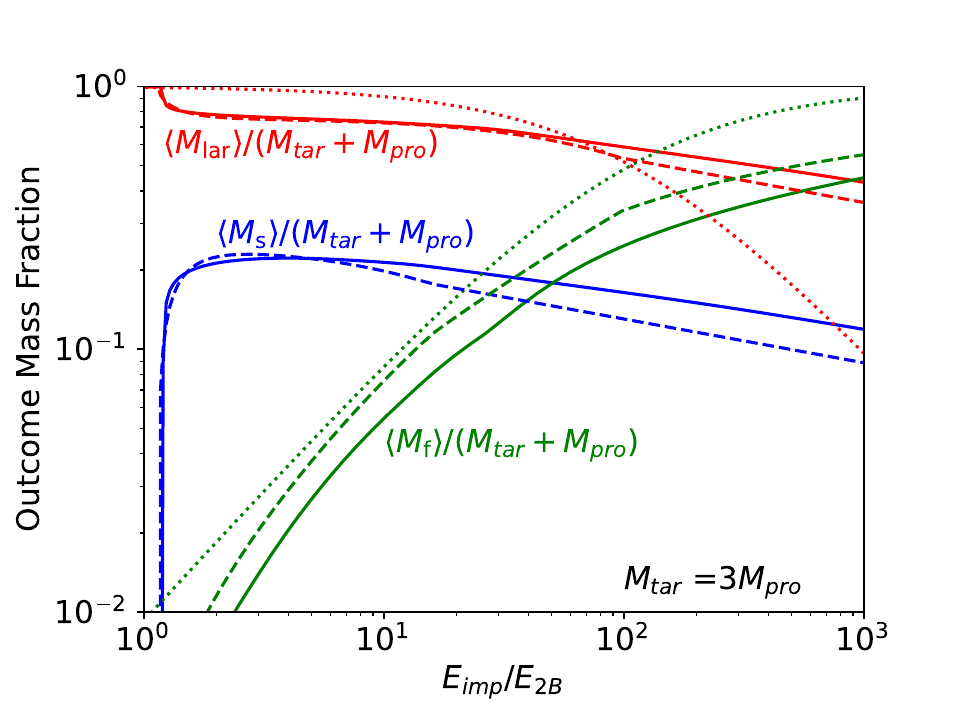}{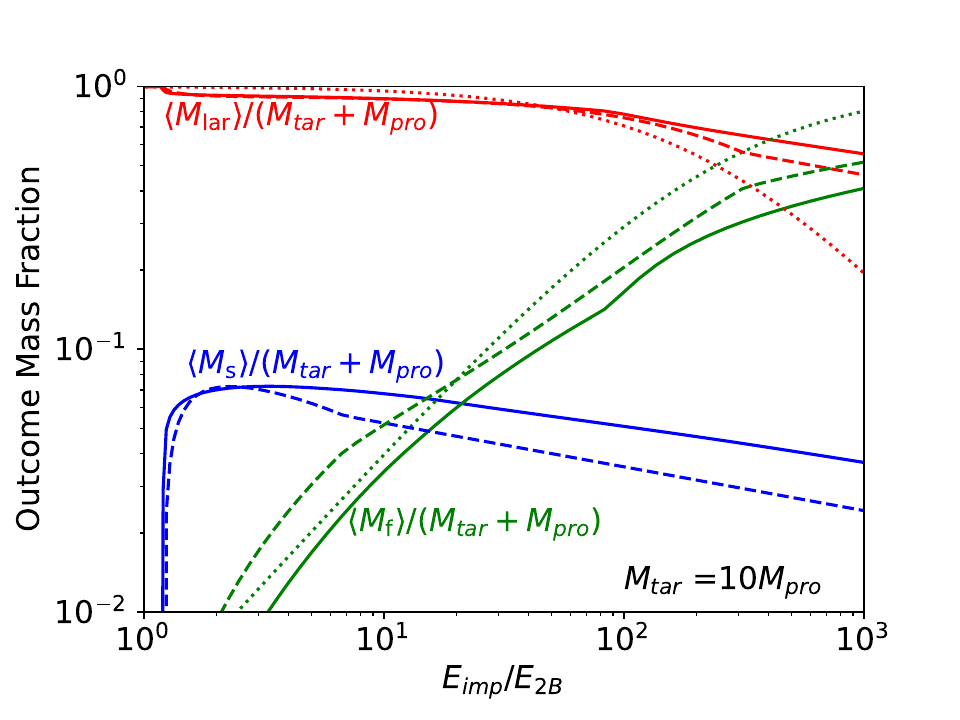}
 \figcaption{Impact-parameter-weighted average collisional outcomes with the analytic formulae of Eqs.~(\ref{eq:ana_mlar}) and (\ref{eq:ana_ms}) for $M_{\rm tar} / M_{\rm pro} = 3$ (left) and 10 (right): largest mass $\langle M_{\rm lar} \rangle/(M_{\rm tar}+M_{\rm pro})$ (red), second largest mass $\langle M_{\rm s} \rangle/(M_{\rm tar}+M_{\rm pro})$ (blue), and fragment mass $\langle M_{\rm f} \rangle/(M_{\rm tar}+M_{\rm pro})$ (green), each normalized by the total mass of the colliding bodies. Solid lines show results from the formulae in \S~\ref{sc:formula}. Dashed lines indicate the new approximate formulae given in Eqs.(\ref{eq:ave_ml}), (\ref{eq:ave_ms}), and (\ref{eq:ave_mf}). Dotted lines represent the previous formula by Eq.~(\ref{eq:previous_model}).
\label{fig:average}}
\end{figure*}

%fof/outcome_fomula.py
The ``impact parameter'' weighted average for collisional outcomes are practically needed in numerical simulations of collisional growth toward planet formation. 
\rev{
Isotropic collisions provide a reasonable approximation in planet formation, although 
minor nonuniformities arising from gravitational interaction are important for generating planetary rotation by planetesimal accretion \citep{ohtsuki98}. Under the assumption of isotropic collisions, }
the average of $M_{\rm lar}$ is given by 
\begin{equation}
 \langle M_{\rm lar} \rangle = \int_0^{\pi/2} M_{\rm lar} \sin \theta d \theta, 
\end{equation}
and $\langle M_{\rm s} \rangle$ is calculated similarly. 

Figure \ref{fig:average} shows $\langle M_{\rm lar} \rangle$, $\langle M_{\rm s} \rangle$, and $\langle M_{\rm f} \rangle$ using the analytic outcome formulae given in \S~\ref{sc:formula}, which are consistent with the average outcomes of simulations shown in \S.~\ref{sc:simulation}. 
%For high $E_{\rm imp}$, 
For $E_{\rm imp} / E_{\rm 2B}\leq 1.2$ -- 1.3, $\langle M_{\rm lar} \rangle \approx M_{\rm tar} + M_{\rm pro}$ and $\langle M_{\rm s} \rangle \ll M_{\rm tar} + M_{\rm pro}$, which are mainly determined by the ``merging'' collisions. 
For $E_{\rm imp}/E_{\rm 2B} \ga 100$, $\langle M_{\rm lar} \rangle$ and $\langle M_{\rm s} \rangle$ monotonically decrease with $E_{\rm imp}$. The remnant masses $M_{\rm lar}$ and $M_{\rm s}$ are inversely proportional to $E_{\rm imp}$ for high $E_{\rm imp}$ unless vaporization effectively occurs for ${\tilde E}_{\rm v,tot} \ll 1$ (see the formulae in \S~\ref{sc:formula}). However, average masses $\langle M_{\rm lar} \rangle$ and $\langle M_{\rm s} \rangle$ decreases more slowly with $E_{\rm imp}$. For high angle collisions, the significant mass reduction of $M_{\rm lar}$ and $M_{\rm s}$ requires very high $E_{\rm imp}$, while low-angle collisions result in small $M_{\rm lar}$ and $M_{\rm s}$ even for low $E_{\rm imp}$ (see Figures \ref{fig:comp_ml} and \ref{fig:comp_second}). The angle dependence gives the slow dependence of average outcomes on $E_{\rm imp}$. 

We have investigated the effect of collisional fragmentation in planet formation \citep{kobayashi+10,kobayashi12,kobayashi13,kobayashi18}, where we have modeled $M_{\rm lar}$ as \citep{kobayashi_tanaka10,kobayashi+10}, 
\begin{eqnarray}
 \langle M_{\rm lar} \rangle &=& \frac{M_{\rm tar}+M_{\rm pro}}{1+\phi},\label{eq:previous_model} \\
 \phi &=& \frac{E_{\rm imp}}{(M_{\rm tar}+M_{\rm pro})Q_{\rm D}^\star}, 
\end{eqnarray}
where $Q_{\rm D}^\star$ is the specific energy for disruption. We plot this formula with $Q_{\rm D} = 10 v_{\rm esc}^2$ in Figure \ref{fig:average}. Since $M_{\rm s}$ is not defined in the model, we plot $M_{\rm tar} + M_{\rm pro} - \langle M_{\rm lar} \rangle$ instead of $\langle M_{\rm f} \rangle$ (green dotted lines in Figure~\ref{fig:average}). From the comparison with the outcome model presented in this paper, the previous model overestimates collisional fragmentation for high $E_{\rm imp}$. This is caused by the outcomes for high $\theta$, as discussed above. 
In addition, the current model includes the effect of hit-and-run collisions, which hinders for simple collisional growth at $E_{\rm imp}/E_{\rm 2B} \sim 10$. 

The average outcome formulae cannot be derived analytically from the formulae in \S \ref{sc:formula}. 
We approximately provide the average outcome formulae.
The formulae for $\langle M_{\rm lar} \rangle$ and $\langle M_{\rm s} \rangle$ are following. 
\begin{equation}
\langle M_{\rm lar} \rangle = \left\{
\begin{array}{l}
(M_{\rm tar} + M_{\rm pro}) / (1+ E_{\rm imp}/E_{\rm mod})\\
\quad \quad\quad{\rm for}\, E_{\rm imp} < E_{\rm hr}, \\
%1.2/(1-\gamma^\prime_{\rm a}),\\
%
M_{\rm l,a,s} + M_{\rm l,tra}
%(1-\gamma^\prime_{\rm a}) {\tilde M}_{\rm s,a,s} 
%\frac{E_{\rm imp}}{E_{\rm 2B}},   
\\ \quad \quad \quad {\rm for}\, 
%1.2 / (1-\gamma^\prime_{\rm a}) \leq E_{\rm imp}/ E_{\rm 2B} < 1/5 \psi_1, 
E_{\rm hr} \leq E_{\rm imp} < E_{\rm l,ero}, 
\\
\, {\rm Max}(M_{\rm l,a,s},M_{\rm l,a,h})
+ M_{\rm l,tra}
%M_{\rm pro} 
%(1-\gamma^\prime_{\rm a}) {\tilde M}_{\rm s,a,s} 
%\frac{E_{\rm imp}}{E_{\rm 2B}}.
\\ \quad \quad\quad{\rm for}\, 
%E_{\rm imp}/ E_{\rm 2B} \geq 1/5 \psi_1, 
E_{\rm imp} \geq E_{\rm l,ero}, 
\end{array}
\right.\label{eq:ave_ml}
\end{equation}
\begin{equation}
\langle M_{\rm s} \rangle = \left\{
\begin{array}{l}
0.01 M_{\rm tar} / (1+ E_{\rm mod}/E_{\rm imp})\\
\quad \quad\quad{\rm for}\,  E_{\rm imp} < E_{\rm hr}, 
\\
   M_{\rm s,a,s} - M_{\rm s,tra}
\\
\quad \quad\quad{\rm for}\,  E_{\rm hr} \leq E_{\rm imp} < E_{\rm s,ero}, 
\\
   {\rm Max}(M_{\rm s,a,s},M_{\rm s,a,h})- M_{\rm s,tra}
\\ \quad \quad\quad{\rm for}\, E_{\rm imp} > E_{\rm s,ero}, , 
\end{array}
\right.\label{eq:ave_ms}
\end{equation}
where 
\begin{equation}
E_{\rm hr} = \frac{E_{\rm 2B}}{\psi_0},
%1-\gamma_{\rm a}^\prime},
\quad
  E_{\rm l,ero} = \frac{E_{\rm 2B}}{5 \psi_1}, \quad
  E_{\rm s,ero} = \frac{E_{\rm 2B}}{5 \psi_2}, 
\end{equation}
\begin{eqnarray}
% E_{\rm mod} &=& 0.25 %0.15 \frac{5}{3} 
%%\gamma^\prime_{\rm a} 
%(1-\psi_0)
%E_{\rm imp}
%\nonumber \\
%&& \times
%\frac{(M_{\rm tar}/M_{\rm pro})(1+ M_{\rm tar}/M_{\rm pro})^{-5/3}}{1+(M_{\rm tar}/M_{\rm pro})^{1/3}}
%
E_{\rm mod} &=& 
\frac{4 E_{\rm 2B}}{1-\psi_0}
\left(\frac{M_{\rm tar}}{M_{\rm pro}}\right)^{-1}
\nonumber
\\
&&\times
\left[
1+\left(
\frac{M_{\rm tar}}{M_{\rm pro}}\right)^{\frac{1}{3}}
\right]
\left(
1+\frac{M_{\rm tar}}{M_{\rm pro}}\right)^{\frac{5}{3}},
%
%  \frac{[1+ M_{\rm tar}/M_{\rm pro})]^2}{1+(M_{\rm tar}/M_{\rm pro})^{1/3}}
%   \left(\frac{M_{\rm tar}}{M_{\rm pro}}\right)^{-2/3}, 
%find the error
\end{eqnarray}
\begin{flalign}
& M_{\rm l,a,s} = %\frac{1}{1+E_{\rm imp}/5E_{\rm l,ero}}, 
\left(1+\frac{E_{\rm imp}}{5E_{\rm l,ero}}\right)^{-1} M_{\rm tar}, &
\end{flalign}
\begin{flalign}
& M_{\rm l,a,h} = 1.51 %1.4 % sum up the pre-factors 
%(0.15*3/16)**(-1/6)*(1/5/0.18**(5/6))= 1.5139..
\left[
\frac{E_{\rm imp}}{E_{\rm 2B}}\frac{(M_{\rm tar}/M_{\rm pro})^2+1}{(M_{\rm tar}/M_{\rm pro})+1} \right]^{-\frac{1}{6}}
&
\nonumber
\\
& \quad \quad \quad \quad \times 
%&& \times 
\left(\frac{M_{\rm tar}}{M_{\rm pro}}\right)^{\frac{2}{9}} M_{\rm tar},&
\end{flalign}
\begin{flalign}
&M_{\rm l,tra} = \left[\left(M_{\rm pro}-M_{\rm s,a,s} \right) \left(\frac{E_{\rm hr}}{E_{\rm imp}}\right)
+M_{\rm s,tra}\right], &
\end{flalign}
\begin{flalign}
&M_{\rm s,tra} = 
 \left(1 - \sqrt{1- \frac{E_{\rm hr}^4}{E_{\rm imp}^4}} \right) M_{\rm s,a,s},  &
\end{flalign}
\begin{flalign}
 & M_{\rm s,a,s} = \left(1+\frac{E_{\rm imp}}{5E_{\rm s,ero}}\right)^{-1} M_{\rm pro}, &\\
 & M_{\rm s,a,h} = 0.84 % sum up the pre-factors 
\left(
%\psi_2 
\frac{E_{\rm imp}}{5E_{\rm s,ero}}\right)^{-\frac{1}{6}} M_{\rm pro},&
\end{flalign}
using 
\begin{eqnarray}
% \gamma^\prime_{\rm a} 
\psi_0
&=& \frac{2}{3} + \frac{2}{9}\exp\left(-\frac{[\log_{10}(M_{\rm tar}/M_{\rm pro})]^2}{3}\right),\\
 \psi_1 &=& \frac{1}{6} %0.1*5/3
%\gamma^\prime_{\rm a} 
(1-\psi_0)
\frac{(M_{\rm tar}/M_{\rm pro})^{-2/3}}{(M_{\rm tar}/M_{\rm pro})^{1/3}+1}, \\
 \psi_2 &=& \frac{1}{6} %0.1 \frac{5}{3}
%\gamma^\prime_{\rm a}
(1-\psi_0)
\frac{(M_{\rm tar}/M_{\rm pro})}{(M_{\rm tar}/M_{\rm pro})^{1/3}+1}.
\end{eqnarray}
In addition, 
\begin{equation}
 \langle M_{\rm f}\rangle = M_{\rm tar} + M_{\rm pro} - \langle M_{\rm lar}\rangle - \langle M_{\rm s}\rangle.\label{eq:ave_mf} 
\end{equation}

The formulae in Eqs.~(\ref{eq:ave_ml}), (\ref{eq:ave_ms}), and (\ref{eq:ave_mf}) are approximately in agreement with the analytic formulae (see Fig.~\ref{fig:average}). 

\section{Discussion}
\label{sc:discussion}

\subsection{Characteristic Impact Energy}
\label{sc:character}

\begin{figure*}[hbt]
\plottwo{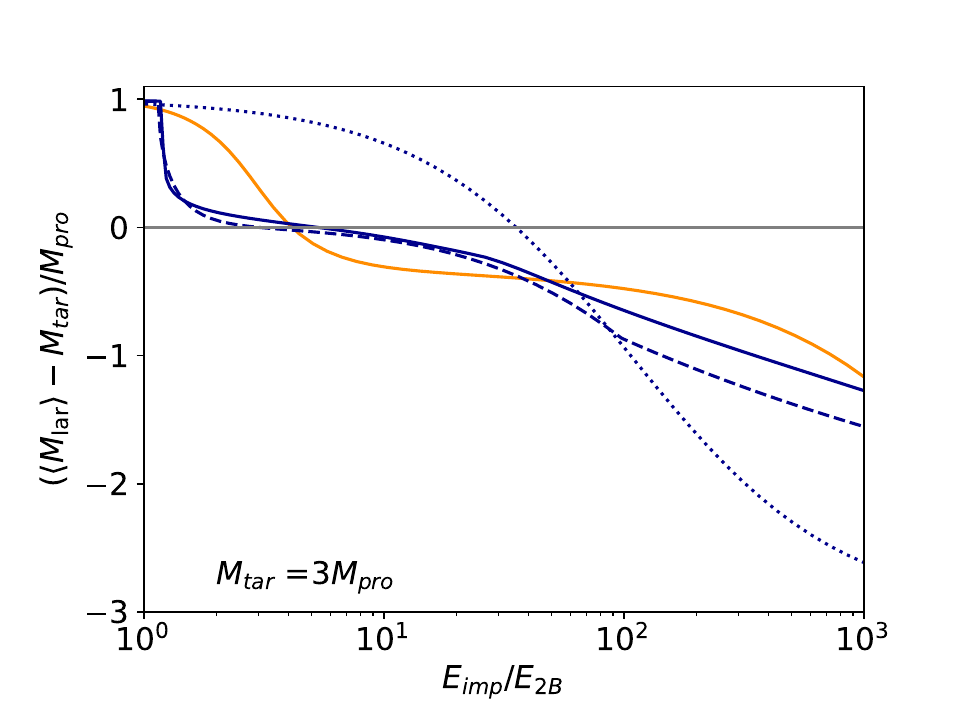}{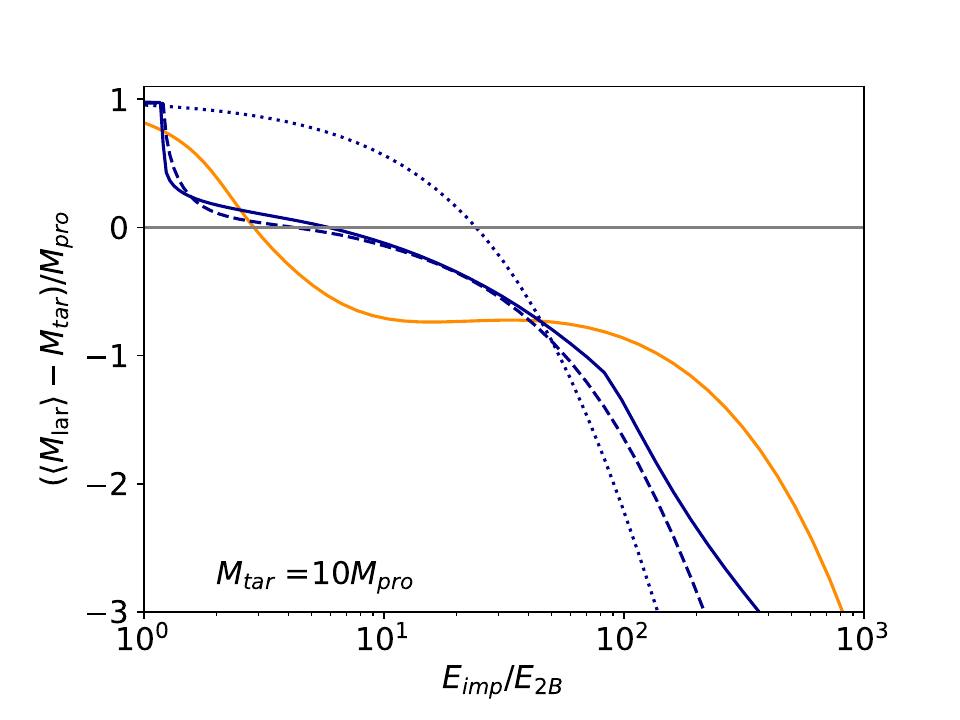}
\figcaption{Growth efficiency $(\langle M_{\rm lar} \rangle - M_{\rm tar})/M_{\rm pro}$ as a function of $E_{\rm imp}/E_{\rm 2B}$. Solid, dashed, and dotted curves represent the same as in Fig.~\ref{fig:average}, but plotted for the growth efficiency. The orange curves represent the fitting formula derived from dust aggregate collisional simulations by \citet{hasegawa23}, where the horizontal axis is given by $E_{\rm imp} / (M_{\rm pro}+M_{\rm tar}) v_{\rm bre}^2$ instead of $E_{\rm imp}/E_{\rm 2B}$. The horizontal line indicates the threshold where $\langle M_{\rm lar} \rangle = M_{\rm tar}$. 
\label{fig:ave_eff} } 
\end{figure*}

In planet formation, colliding bodies rotate around the host star. The relative velocities in their orbits, $v_{\rm r}$, is related to $v_{\rm imp}$ as
\begin{equation}
 v_{\rm r}^2 = v_{\rm imp}^2 - v_{\rm esc}^2.\label{eq:relative_velocity} 
\end{equation}

The transition between ``merging'' and ``hit-and-run'' determines whether collisions contribute significantly to planetary growth. The average outcome formula in Eq.~(\ref{eq:ave_ml}) give the average transition energy as
\begin{equation}
 E_{\rm imp} = E_{\rm 2 B} / \psi_0 \approx (1.1, \,\dots,\, 1.5)\times E_{\rm 2B}.\label{eq:transition_m_hr} 
\end{equation}
This transition condition is consistent with that empirically obtained from impact simulations by \citet{genda12} if $M_{\rm tar} \approx M_{\rm pro}$. 
From Eqs.~(\ref{eq:relative_velocity}) and (\ref{eq:transition_m_hr}), the transitional relative velocity is given by 
\begin{equation}
 v_{\rm r,t} \approx (0.4, \, \dots, \, 0.7) \times v_{\rm esc}, 
\end{equation}
where $0.4 v_{\rm esc}$ and $0.7 v_{\rm esc}$ is the values for $M_{\rm tar} \gg M_{\rm pro}$ and $M_{\rm tar} = M_{\rm pro}$, respectively. 

Even for ``hit-and-run'', the resulting increase of $\langle M_{\rm lar} \rangle$ is not negligible and  
collisions lead to mass growth if $\langle M_{\rm lar} \rangle > M_{\rm tar}$. The collisional mass loss is characterized by erosive collisions, for which the mass loss of $M_{\rm tar} - \langle M_{\rm lar} \rangle$ is comparable to or larger than $M_{\rm pro}$. To characterize such phenomena, we plot the growth efficiency $(\langle M_{\rm lar} \rangle - M_{\rm tar})/M_{\rm pro}$ in Fig.~\ref{fig:ave_eff}. 
We can easily determine $E_{\rm imp}^*$ at $\langle M_{\rm lar} \rangle = M_{\rm tar}$ from Eq.(\ref{eq:ave_ml}) because $M_{\rm s,tra} \ll M_{\rm s,a,s}$ for $M_{\rm tar} \gg M_{\rm pro}$. 
\begin{equation}
 \frac{E_{\rm imp}^*}{E_{\rm 2B}} \approx \sqrt{27} \approx 5.2.\label{eq:eg} 
\end{equation}
The estimate is in good agreement with the case with $M_{\rm tar} \sim M_{\rm pro}$ (see Fig.~\ref{fig:vg_comp}). 
The impact velocity corresponding to $E_{\rm imp}^*$ is $v_{\rm imp}^* \approx 2.3 v_{\rm esc}$. 
According to Eqs.~(\ref{eq:relative_velocity}) and (\ref{eq:eg}), the relative velocity at $\langle M_{\rm lar} \rangle = M_{\rm tar}$ is given by 
\begin{equation}
 v_{\rm r,g} \approx 2.0 \, v_{\rm esc}. 
\end{equation}
Therefore, collisional growth still occurs at $v_{\rm r,t} < v_{\rm r} < v_{\rm r,g}$, while collisions with $v_{\rm r} < v_{\rm r,t}$ induces significant growth. 
Such growth may be important for planetesimal growth prior to the onset of runaway growth \citep{kobayashi16}. 

On the other hand, collisional disruption has been investigated well in the field of collisional physics. 
The characteristic energy for collisional disruption
 is defined by $Q_{\rm D}^*$, the specific impact energy at $M_{\rm lar} = M_{\rm tar}/2$. We obtain $Q_{\rm D}^*$ from our analytic formula with $M_{\rm l,a,s} = 1/2$, given by 
\begin{equation}
 Q_{\rm D}^* = (9, \, \dots, \, 14)\times v_{\rm esc}^2, \label{eq:qds} 
\end{equation}
where the values are $9 v_{\rm esc}^2$ for $M_{\rm tar} \gg M_{\rm pro}$ and $14 v_{\rm esc}^2$ for $M_{\rm tar} \sim M_{\rm pro}$. 
The result is consistent with $Q_{\rm D}^*$ in the gravity dominated regime \citep{stewart09}. 
Interestingly, $Q_{\rm D}^*$ and $v_{\rm imp}^*$ are independent of $M_{\rm tar}/M_{\rm pro}$, if we ignore the weak dependence of $v_{\rm esc}$ on $M_{\rm tar}/M_{\rm pro}$. 
%Therefore, collisional growth no longer occurs, if bodies have impact velocities larger than $v_{\rm imp}^*$. 
It should be noted that this result comes from the simple assumption of the self-gravity and shock dominance, which is valid only for 1,000\,km-sized or larger planetesimals. In addition, such high-velocity collisions result in high internal energies. Pressure is mainly determined by the thermal term (the first term in the right hand side of Eq.~8). The impact shock then directly leads to the energy dissipation as we consider in \S~\ref{sc:formula}. However, the cold pressure, the second and third terms in the right hand side of Eq.~(8) is more important for small planetesimal impacts, which results in the velocity dependence of $Q_{\rm D}^*$ \citep{suetsugu18}. 
In addition, compaction, friction, and cohesion enhance $Q_{\rm D}^*$ of small planetesimals compared with Eq.~(\ref{eq:qds}) \citep{jutzi15}. 

For 1,000\,km-sized or larger planetesimals, the characteristic relative velocities $v_{\rm r,t}$ and $v_{\rm g,t}$ and $Q_{\rm D}^*$ have been obtained from the new formula, which are important for collisional growth or fragmentation of planetesimals. 
For smaller planetesimals, energy dissipation in addition to shock is effective, which may lead to the enhancement of $v_{\rm r,t}$ and $v_{\rm r,g}$ as well as $Q_{\rm D}^*$.

\subsection{Comparison with Dust Aggregate Collisions \label{sc:dust_model}}

Early planet formation occurs via collisional agglomeration of dust particles in protoplanetary disks. Collisions between dust aggregates result in the similar collisional outcomes: ``merging'', ``hit-and-run'', and ``fragmentation'' \citep{hasegawa21}. Therefore, we here compare the formula to dust aggregate collisions. 

In the new formula, the collisional outcome is given by a function of $E_{\rm imp}/E_{\rm 2B}$. For dust aggregates, $E_{\rm 2B}$ is negligible, while cohesion is important. We adopt $(M_{\rm tar}+M_{\rm pro}) v_{\rm bre}^2$ instead of $E_{\rm 2B}$ in the new formula, where $v_{\rm bre}$ is the breaking velocity of a single contact between two monomers \citep{hasegawa23}.

\begin{figure}[hbt]
\plotone{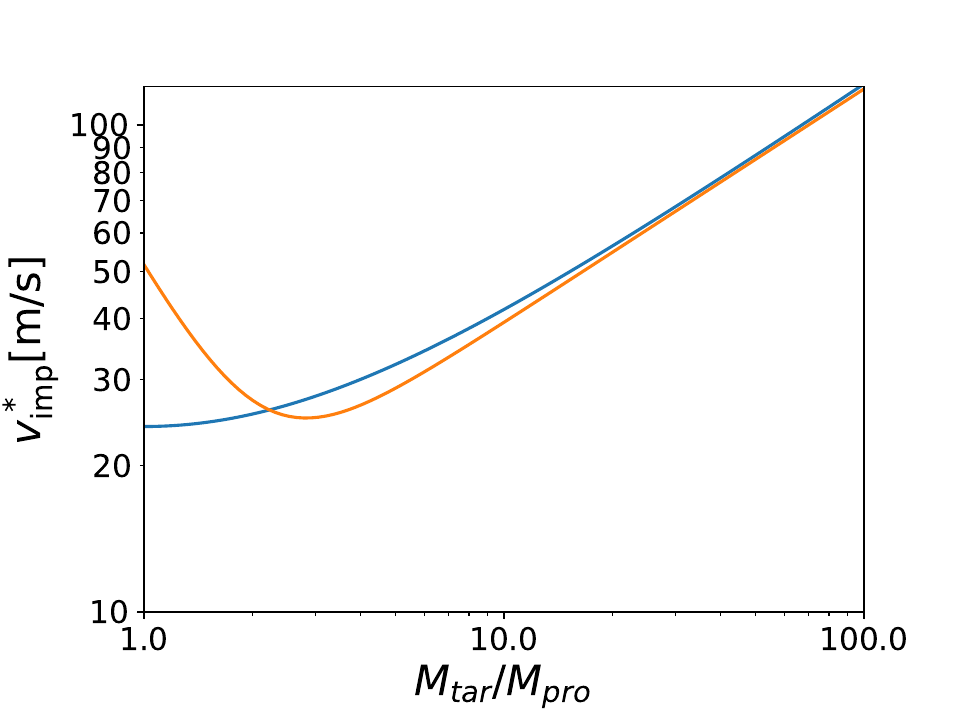}
\figcaption{Dust impact velocity at $\langle M_{\rm lar} \rangle = M_{\rm tar}$, $v_{\rm imp}^*$, as a function of $M_{\rm tar}/M_{\rm imp}$. 
The blue curve represents the new formula with $(M_{\rm pro}+M_{\rm tar}) v_{\rm bre}^2$, while the orange curve corresponds to the fitting formula derived from dust aggregate simulations in \citet{hasegawa21}. 
\label{fig:vg_comp} } 
\end{figure}

We then compare the new formula with the result of dust aggregate collisions. 
Figure \ref{fig:vg_comp} shows the impact velocity at $\langle M_{\rm lar} \rangle = M_{\rm tar}$, $v_{\rm imp}^*$, for dust aggregate impact simulations and the new formula with $(M_{\rm tar}+M_{\rm pro}) v_{\rm bre}^2$ instead of $E_{\rm 2B}$. The modified new formula 
well reproduces the result of simulation for $M_{\rm tar} \gg M_{\rm pro}$, while $v_{\rm imp}^*$ is underestimaed for $M_{\rm tar} < 2 M_{\rm pro}$. 
Even for the crude assumption of $(M_{\rm tar}+M_{\rm pro}) v_{\rm bre}^2$ instead of $E_{\rm 2B}$, 
the new formula works well. 

We compare the dependence of $(\langle M_{\rm lar} \rangle - M_{\rm tar})/M_{\rm pro}$ on $E_{\rm imp}$ (see Fig.~\ref{fig:ave_eff}). The new formula approximately reproduces the general collisional regimes: ``merging'' for low energies, ``hit-and-run'' for intermediate energies, and ``catastrophic disruption'' for high energies. However, further refinement is required to accurately capture the detailed outcomes of dust aggregate collisions.

\subsection{Implication of the New Model to Planet Formation}

In the planetesimal accretion of protoplanets, collisional fragmentation of planetesimals is caused by the stirring by protoplanets \citep{wetherill89,inaba01}.
Planetesimals are reduced by radial drift of small bodies produced by the collisional cascade induced by planetesimal fragmentation \citep{kobayashi+10,kobayashi11,kobayashi18}. The depletion timescale of planetesimals due to the collisional cascade is estimated from the formula by \citet{kobayashi_tanaka10}. We adopt the new collisional outcome given by Eq. (\ref{eq:ave_ml}) in the collisional cascade formula. A timescale for the collisional cascade with the new outcome model is roughly comparable to that with the old collisional outcome model. Therefore, the new collisional model insignificantly influences planet growth by the planetesimal accretion. 

Recently, the direct simulation of collisional evolution from dust to planets is possible \citep{kobayashi21}. The collisional growth of dust aggregates produces planetesimals at $\la 10$\,au in protoplanetary disks with usual temeratures, which forms the solid cores of gas giant planets in $\sim 10^5$ years \citep{kobayashi23,ikoma_kobayashi25}. For such simulations, a collisional model is required to connect dust to planets. According to the consideration in \S~\ref{sc:dust_model}, we need to expand the new collisional outcome formula for the direct simulation from dust to planets. 

After the significant depletion of protoplanetary disks, the long-term chaotic orbital evolution of protoplanets occurs \citep{iwasaki01,iwasaki02,kominami02,agnor02}. In the later stage, collisions between protoplanets are a dominant process for the growth of planet masses. The new model includes ``hit-and-run'' and ``erosive'' collisions, which frequently occur and produce small mass ejecta in this stage. 
\citet{genda15b} showed the total ejecta mass in a series of giant impacts is comparable to an Eath mass, via SPH impact simulations using impact parameters obtained from $N$-body simulations of protoplanets without collisional fragmentation. \citet{izidoro22} carried out $N$-body simulations with the outcome model by \citet{leinhardt12}, without modeling the collisional evolution between giant impact ejecta.  
\citet{kobayashi19} showed that collisional evolution between small bodies surrounding around protoplanets effectively decreases the total mass of surrounding bodies, and giant impact ejecta may influence the orbital evolution of protoplanets if largest ejecta is larger than 1000\,km in radius. 
The present new model provides the collisional outcome of protoplanets and their outcome velocities using the dissipation of $E_{\rm over,\perp}$ for $N$-body simulations of protoplanets. 
Therefore, 
our findings may provide a step toward improved descriptions of planetary growth in the later stages.

\section{Summary}

We have developed a new analytic model for collisional outcomes of 1,000km-sized or larger colliding bodies, calibrated with SPH impact simulations. 
The model incorporates characteristic energies---particularly those associated with overlapping masses and perpendicular impact energy $E_{\rm over,\perp}$ (see Eq.~\ref{eq:eoperp})---to reproduce transitions among merging, hit-and-run, and catastrophic disruption. 

\begin{itemize}
 \item \textbf{Characteristic impact energy:} 
       The transition is determined by the residual energy, $E_{\rm res} = E_{\rm imp} - E_{\rm over,\perp}$, in comparison with the two-body gravitational binding energy $E_{\rm 2B}$ (see Eq.~\ref{eq:e_2b}). If $E_{\rm res} \ll E_{\rm 2B}$, the colliding bodies merge, whereas $E_{\rm res} \ga E_{\rm 2B}$ results in ``hit-and-run'' outcomes. 

% \item \textbf{Collisional growth and fragmentation:} 
       The transition between ``merging'' and ``hit-and-run'' occurs at the impact energy of $(1.1,\, \dots, \, 1.5) \times E_{\rm 2B}$ (see Eq.~\ref{eq:transition_m_hr}). Significant collisional growth occurs in the ``merging'' regime. Even for ``hit-and-run'' collisions, the collisional growth occurs slowly. The threshold for mass growth occurs at an impact energy of $\sim 5 E_{\rm 2B}$ (see Eq.~\ref{eq:eg}). For significant collisional fragmentation by disruptive disruption, the specific disruption energy is $(9,\, \dots,\, 14) \times v_{\rm esc}^2$ (see Eq.~\ref{eq:qds}). The collisional evolution is characterized by the three energies. 
    
 \item \textbf{New collisional outcome formula:} We provide the new collisional outcome formula, which expresses the characteristic energies properly. The formula in Eq.~(\ref{eq:ana_mlar}) and (\ref{eq:ana_ms}) is useful for $N$-body simulations in planet formation. The impact-parameter-average formula in Eqs.~(\ref{eq:ave_ml}) and (\ref{eq:ave_ms}) are applicable for the statistical approach simulation of planetesimal accretion. 

 \item \textbf{Dust aggregates:} 
       By replacing gravitational binding energy with contact-breaking energy,
       the model also reproduces collisional regimes observed in dust aggregate
       simulations, linking small-scale and large-scale collisions (see \S.~\ref{sc:dust_model}). 
\end{itemize}

Overall, the new outcome formula provides a unified framework for collisional processes across a wide range of body sizes---from dust aggregates to planetary embryos---and offers an improved basis for modeling planet formation.

\appendix
\section{Tillotson equation of state}
\label{sc:tillotson}

If $\rho \geq \rho_0$ or $u \leq u_{\rm iv}$, the form is described by Eq.~(\ref{eq:tillotson}), which is for the solid state. For $\rho < \rho_0$ and 
$u > u_{\rm cv}$, the form is given by 
\begin{eqnarray}
 P &=& a \rho u  
+ \left[\frac{b \rho u}{u \rho_0^2/ u_0 \rho^2 + 1} +A \frac{\rho - \rho_0}{\rho_0} e^{-\beta_{\rm T}(\rho_0/\rho-1)}\right]
\nonumber\label{eq:pressure_vapor}
\\  
&& \times
e^{-\alpha_{\rm T}(\rho_0/\rho-1)^2},
\end{eqnarray}
which expresses the complete vaporization. For $\rho < \rho_0$ and $u_{\rm iv} < u < u_{\rm cv}$, the form is given by 
\begin{equation}
 P = \frac{(u-u_{\rm iv})P_{\rm E} + (u_{\rm cv}-u)P_{\rm C}}{u_{\rm cv} - u_{\rm iv}}, 
\end{equation}
where $P_{\rm E}$ and $P_{\rm C}$ are the pressures calculated from Eq.~(\ref{eq:pressure_vapor}) and (\ref{eq:tillotson}). It expresses the state of partial vaporization. 

\section{overlapping mass}
\label{sc:overlap}

\begin{figure}
 \plotone{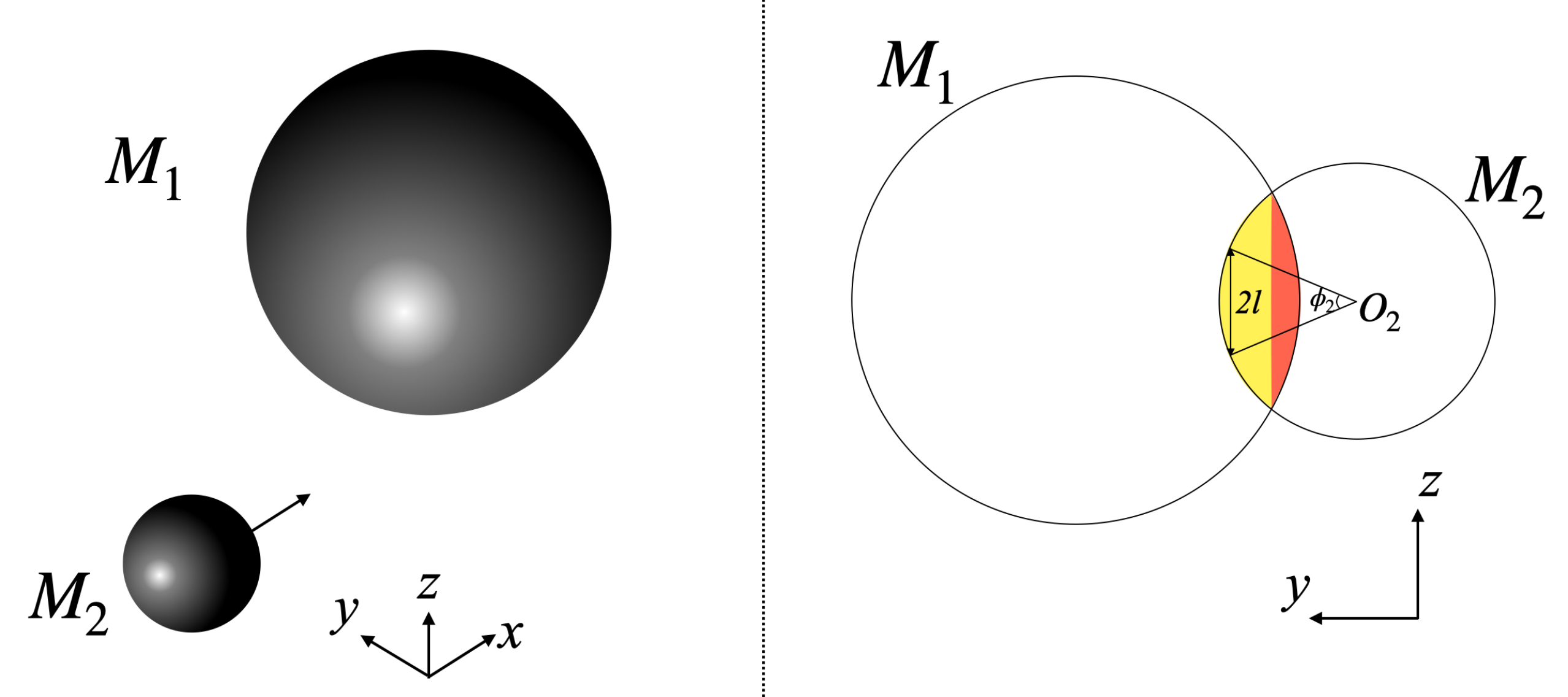}
 \figcaption{Schematic of overlapping volumes. (Left) Impact geometry in the $x$-$y$-$z$ coordinate system. (Right) Overlapping region (yellow and orange) projected onto the $y$-$z$ plane.
\label{fig:overlap}
}
\end{figure}

We consider an impact between objects 1 and 2 with radii $R_1$ and $R_2$, respectively. 
The overlap volume during an impact for the object 1 (the target) 
is estimated as the sum of two volumes. 
\begin{equation}
 V_{\rm o,1} = V_{{\rm in},1} + V_{\rm out,1}. 
\end{equation}
$V_{{\rm in},1}$ is the volume integrated from the yellow area in Figure \ref{fig:overlap}, while $V_{{\rm out},1}$ is the volume integrated from the orange area in Figure \ref{fig:overlap}. 

The volume $V_{{\rm in},1}$ is obtained from 
\begin{equation}
 V_{{\rm in},1} = \int_{\xi_2}^1 S(\xi) R_2 d\xi, 
\end{equation}
where 
\begin{equation}
 \xi_2 = {\rm Max}\left(\frac{R_{\rm 2}^2 + R_{12}^2 - R_1^2}{2 R_{\rm 2} R_{12}}, -1 \right), 
\end{equation}
and the area $S(\xi)$ given by 
\begin{equation}
 S(\xi) = 2 l(\xi) \sqrt{r(\xi)^2 - l(\xi)^2} + 2r(\xi)^2 \arcsin
 \left(\frac{l(\xi)}{r(\xi)}\right),
\end{equation}
$r^2 = R_{\rm 1}^2 - (R_{12} - R_2 \xi)^2$, $l^2 = R_2^2
(1-\xi^2)$, and $R_{12} = (R_1 + R_2) \sin \theta$. 
The length $l$ at $\xi = \cos \phi_2$ is illustrated in Figure \ref{fig:overlap}. 
% \begin{equation}
%  S(\xi) = \left\{
% \begin{array}{c c}
% 2 l(\xi) \sqrt{r(\xi)^2 - l(\xi)^2} + 2r(\xi)^2 \arcsin
% \left(\frac{l(\xi)}{r(\xi)}\right)
% & {\rm if \,} l < r, \\
% \pi r(\xi)^2 & {\rm otherwise,}%{\rm if \,} l \geq r, 
% \end{array}
% \right.
% \end{equation}
%where $r^2 = R_{\rm 1}^2 - (R_{12} - R_2 \xi)^2$ and $l^2 = R_2^2
%(1-\xi^2)$ with being $R_{12} = (R_1 + R_2) \sin \theta$. 

The volume $V_{\rm out,1}$ is given by 
\begin{equation}
 V_{\rm out,1} = 
\frac{\pi R_1^3}{3} (2 - 3 \xi_1 + \xi_1^3),
\end{equation}
where 
\begin{equation}
\displaystyle
 \xi_1 = {\rm Min}(\frac{R_{\rm 1}^2 + R_{12}^2 - R_2^2}{2 R_{\rm 1} R_{12}},1).
\end{equation}

The overlapping volumes in cores, $V_{\rm o,c,1}$ and $V_{\rm o,c,2}$ are calculated from the same equation using the core radii $R_{\rm c,1}$ and $R_{\rm c,2}$ instead of $R_1$ and $R_2$, respectively. We then calculate the overlapping mass from the overlapping volumes and the average densities of the core and mantle.

\begin{acknowledgements}
\rev{We thank the anonymous referee for the careful review and useful comments. }
The work is supported by Grants-in-Aid for Scientific Research
(JP21K03642, JP22H01278, JP22H00179, JP24K00690, JP25K01055) from MEXT of Japan.  Numerical
computations were carried out on Cray XC50 and HPE Cray XD2000 at the Center for
Computational Astrophysics, National Astronomical Observatory of Japan.
\end{acknowledgements}

\bibliography{apj}

\end{document}